\documentclass[journal,oneside,final,twocolumn,letterpaper]{IEEEtran}
\usepackage{amssymb}
\usepackage{graphicx}
\usepackage[english]{babel}
\usepackage{mathrsfs}
\usepackage[tbtags]{amsmath}
\usepackage{amsfonts}
\usepackage{amssymb}
\usepackage{mathrsfs}

\newtheorem{thm}{Theorem}

\newtheorem{lemma}[thm]{Lemma}

\begin{document}
\title{An Asymptotically Optimal Push-Pull Method for Multicasting over a Random Network}

\author {Vasuki~Narasimha~Swamy, Srikrishna~Bhashyam,~\IEEEmembership{Senior Member,~IEEE}, Rajesh Sundaresan,~\IEEEmembership{Senior Member,~IEEE}, and Pramod~Viswanath,~\IEEEmembership{Fellow,~IEEE}
\thanks{This paper was presented in part at the 2012 IEEE International Symposium on Information Theory (ISIT 2012) held at Cambridge, MA, USA.}
\thanks{Vasuki Narasimha Swamy is with the Department of EECS, University of California at Berkeley, CA, USA. Srikrishna Bhashyam is with the Department of Electrical Engineering at the Indian Institute of Technology Madras, Chennai, India. Rajesh Sundaresan is currently visiting the Coordinated Science Laboratory, University of Illinois at Urbana-Champaign, IL, USA, and on leave of absence from the ECE Department of the Indian Institute of Science, Bangalore, India. Pramod Viswanath is with the Department of ECE and the Coordinated Science Laboratory, University of Illinois at Urbana-Champaign, IL, USA.}
\thanks{This work was supported by the Department of Science and Technology, Government of India, by the University Grants Commission, India, by a fellowship awarded by the Indo-US Science and Technology Forum, and by the US National Science Foundation under grant CCF-1017430.}
\thanks{Parts of this work were carried out when (1) the first author was a student intern and the last author was on sabbatical leave at the Indian Institute of Science, (2) when the first author returned to complete her Bachelors project at the Indian Institute of Technology Madras, and (3) when the third author was on sabbatical leave at the University of Illinois at Urbana-Champaign. Supports from all these host institutions are gratefully acknowledged.}
\thanks{Copyright (c) 2012 IEEE. Personal use of this material is permitted. However, permission to use this material for any other purposes must be obtained from the IEEE by sending a request to pubs-permissions$@$ieee.org.}
}

\maketitle
\begin{abstract}
We consider allcast and multicast flow problems where either all of the nodes or only a subset of the nodes may be in session. Traffic from each node in the session has to be sent to every other node in the session. If the session does not consist of all the nodes, the remaining nodes act as relays. The nodes are connected by undirected links whose capacities are independent and identically distributed random variables. We study the asymptotics of the capacity  region (with network coding) in the limit of a large number of nodes, and show that the normalized sum rate converges to a constant almost surely. We then provide a decentralized push-pull algorithm that asymptotically achieves this normalized sum rate without network coding.
\end{abstract}


\section{Introduction}
\label{sec:introduction}

In this paper, we investigate the capacity of allcast and multicast sessions over random link-capacitated graphs. Two questions motivated us to study these problems in the context of random graphs.

(1) While it is known that network coding in general provides a large coding advantage over multicast flows in directed graphs, Li et al. \cite{Li-et-al} showed that the coding advantage in undirected graphs is upper bounded by 2. In some specific topologies a tighter upper bound is known \cite{201202TIT_MahLiLi}. However several simulation experiments showed nearly no coding advantage for some class of random undirected graphs \cite{Li-et-al-2005}. Is there a provable statement that there is negligible multicast coding advantage for a rich class of random undirected networks?

(2) If we stick to the domain of flows (with duplication), as we will soon see, optimal allcasting and multicasting lead to tree and Steiner tree packing problems respectively. While packing of trees is known to be easy (see \cite{Tutte}, \cite{Nash-Williams}, \cite{Barahona}), Steiner tree packing is known to be hard \cite{200301SODA_JaiMahSal}. Due to its application in multicasting over wired networks and in VLSI layout optimization, practitioners and theorists have over many years provided hardness results, heuristics, and approximation algorithms (see \cite{199702MP_GroMarWei}, \cite{200006ToN_CheGunYen}, \cite{200301SODA_JaiMahSal}, \cite{200410FOCS_Lau}, \cite{200804JCO_SaaEtAl}, etc.) Are there ``quick-but-dirty'' (terminology from \cite{Karp-et-al}), decentralized, scalable, yet near-optimal algorithms for allcasting and multicasting over a rich class of random undirected networks? An answer to this question is of obvious value in the context of live streaming of popular events to a large audience\footnote{On 14 October 2012, an Austrian skydiver Felix Baumgartner broke an existing record for the highest skydive; there were more than 8 million concurrent livestreams of this event on the YouTube video distribution service.}.

In this paper, we provide affirmative answers to both these questions. We begin by making precise what we mean by allcast and multicast.

{\em Allcast}: Consider a setting where there are $n$ nodes, all of which are engaged in a conference over a wired network. Each node has data that needs to be made entirely available over the network to each of the other $n-1$ nodes in a simultaneous fashion. (To be more precise, this is a {\em multiple} allcast problem). The data can be split, or routed, or coded, or transmitted in any combination thereof, so long as all nodes eventually get the information. The underlying complete undirected graph on $n$ vertices is capacitated: each undirected link $e$ has capacity $C_e$ sampled independently and identically from a distribution $F$.  An allcast information flow assignment is said to be feasible if for every link, the net (possibly coded) flow over the link (summed over both directions) respects the link's capacity constraint. For each feasible flow assignment, let $r_i$ be the bit-rate of traffic sent by node $i$ to each of the other nodes. We address the question of the set of all achievable rate tuples $r_1, \cdots, r_n$ in the asymptotics of a large number of nodes $n$. As we shall soon see, this problem is closely related to packing of disjoint spanning trees in a link-capacitated network with integer capacities. Minor extensions of some previous results readily yield that the achievable rate region is almost surely (a.s.)
\begin{equation}
  \label{eqn:all-cast-region}
  \left\{ (r_1, r_2, \ldots) ~:~ \limsup_{n \rightarrow \infty} \frac{1}{n} \sum_{i = 1}^n r_i \leq \frac{1}{2} \mathbb{E}[C] \right\}
\end{equation}
where the expectation is of a random variable $C$ having distribution $F$. The linear programming formulation of this problem is given in Section \ref{sec:LP}, and the proof of (\ref{eqn:all-cast-region}) is given in Sections \ref{sec:upperbound} (converse) and \ref{sec:allcast-achievability} (achievability). Our proof of achievability is via a combination of ``push'' and ``pull'' that suggests a decentralized implementation. Section \ref{sec:isolated-node-no-matching} contains some estimates needed to establish the correctness (with high probability) of the push-pull algorithm. Section \ref{sec:vanishing-pn} deals with the case when the link probabilities vanish, but not too quickly.

It is known that network coding does not yield any coding advantage in allcast settings \cite{Li-et-al}, and thus we have an asymptotic characterization of the allcast capacity region.

{\em Multicast}: We next address a more general setting with only a subset of $k_n$ nodes in the multicast session, where $\lim_{n \rightarrow \infty} k_n / n = \alpha$ and $0 \leq \alpha \leq 1$. Data from each of the $k_n$ nodes has to reach every one of the other $k_n - 1$ nodes. The remaining $n - k_n$ nodes serve as relays. This is therefore a problem of {\em multiple} multicast among common {\em session nodes}. Again, in a link-capacitated framework where each link is independent and identically distributed (iid) with distribution $F$, we are interested in the set of all achievable rate tuples $r_1, \cdots, r_{k_n}$ in the asymptotics of a large number of nodes $n$. We demonstrate that the capacity region is almost surely
\begin{equation}
  \label{eqn:multicast-region}
  \left\{ (r_1, r_2, \ldots) ~:~ \limsup_{n \rightarrow \infty} \frac{1}{n} \sum_{i = 1}^{k_n} r_i \leq \left( 1 - \frac{\alpha}{2}  \right) \mathbb{E}[C]. \right\}
\end{equation}
The LP formulation of this problem is in Section \ref{sec:LP}, proof of the converse is in Section \ref{sec:upperbound}, and proof of achievability is in Section \ref{sec:multicast-achievability}. Here too, our proof of achievability is via a decentralized push-pull algorithm. Section \ref{sec:single-commodity} is a digression to study single commodity flows over random networks and develops the ingredients necessary to establish the correctness (with high probability) of the push-pull algorithm.

Our achievability proofs are based on flows (allowing for duplications) and thus do not employ network coding. In particular, they establish that the coding advantage from network coding in multicast settings, which is the ratio of the maximum achievable rate with network coding and the maximum achievable rate using flows (with duplication), is $1 + o(1)$ as the number of nodes $n \rightarrow \infty$. As the rate achievable without network coding is linear in the number of nodes $n$, the maximum gain to be had from network coding is at best $o(n)$ which is sublinear in the number of nodes. Schemes very similar to our push-pull algorithm have been proposed and are being used over the internet for content distribution in peer-to-peer networks. See \cite[Sec.~1-2]{200504Sigcomm_LiChoZha} for an excellent survey of such techniques. Our work proves that a version of it is asymptotically optimal for a rich class of random networks.

\section{A Linear Programming Formulation}
\label{sec:LP}

\subsection{Random graph models}
\label{subsec:randomness-model}

We are given a countable collection of iid random variables $\{ C_{i,j}, 1 \leq i < j < \infty \}$ where each element has distribution $F$ on $\mathbb{R}_+$. We then obtain a sequence of graphs, denoted $\{ K_n, n \geq 1 \}$, where for each $n$, the graph $K_n$ is the complete graph on the vertex set $\{ 1,2,\ldots,n \}$ along with the collection of all $\binom{n}{2}$ links. Each link $(i,j)$ with $1 \leq i < j \leq n$ has link capacity $C_{i,j}$. Such models are appropriate in settings where nodes are statistically identical in their connections, capacities, and interests. Even in settings where such models are not directly applicable, their tractability yields solutions that provide insights to network designers.

Later on, we will have a need to study Erd\H{o}s-R\'{e}nyi random graphs where the link capacity distribution is Bernoulli$(p)$, which is $\Pr \{C = 1\} = p$ and $\Pr \{ C = 0 \} = 1 - p$. If $C_{i,j} = 0$, then the undirected link $(i,j)$ has zero capacity and is effectively absent. We then use the notation $G(n,p)$ to denote the obtained graph for a fixed $n$.

We will also study Erd\H{o}s-R\'{e}nyi random graphs where $p$ depends on $n$ and vanishes with $n$. We shall denote these $G(n,p_n)$. These may be constructed as follows. We assume that we are now given a collection of iid random variables $\{ Z_{i,j}, 1 \leq i < j < \infty \}$ where each $Z_{i,j}$ has the uniform distribution on $[0,1]$. The graph $G(n,p_n)$ is the graph on $n$ vertices $\{1, 2, \ldots, n \}$ where each link $\{i,j\}$ with $1 \leq i < j \leq n$ has binary capacity $C_{i,j} = {\bf 1} \{ Z_{i,j} \leq p_n \}$. The notation ${\bf 1} \{ \cdots \}$ stands for the indicator of an event. This construction is of course consistent with the construction of $G(n,p)$ when $p_n \equiv p$ is a constant.

Finally, we will also study random bipartite graph sequences $\{ G(n,n,p), n \geq 1 \}$ and $\{ G(n,n,p_n), n \geq 1\}$. These are constructed from the collection of iid random variables $\{ Z_{i,j}, i \geq 1, j \geq 1 \}$ where once again each entry has the uniform distribution on $[0,1]$. In the graph $G(n,n,p_n)$, for example, there are $2n$ vertices with vertex set $V_1 \cup V_2$ where $V_1 = \{ v_1, v_2, \ldots v_n \}$ and $V_2 = \{ \omega_1, \omega_2, \ldots \omega_n \}$, and the capacity on the link between node $v_i$ and node $\omega_j$ is $C_{i,j} = {\bf 1} \{ Z_{i,j} \leq p_n \}$.

\subsection{Allcast}
\label{subsec:LP-allcast}

Consider the allcast problem described in Section \ref{sec:introduction}. Li et al. prove in \cite[Cor. 4.a]{Li-et-al} that a multiple allcast rate vector $(r_1, r_2, \ldots, r_n)$ is achievable in an undirected capacitated network if and only if the rate vector $(\sum_{i=1}^n r_i, 0, \ldots, 0)$ is achievable, i.e., the sum rate is achievable for a {\em single} allcast with node 1 as sender and with the other $n-1$ nodes as receivers. This is intuitively clear since network coding does not help for allcast, and one can make do with multicommodity flows in multiple allcast.

We may therefore assume that there is only one sender (say node 1), and all other $n-1$ nodes are recipients that must receive all information sent by node 1. The rates in such a setting are given by $(r_1, 0, 0, \ldots)$, and we characterize $r_1$.

This maximum rate is obtained by solving the following linear programming (LP) problem. Consider the graph $K_n$ on $n$ vertices with associated link capacities. Let $\mathcal{T}_n$ be the set of all spanning trees on the complete graph (ignoring capacities). The vertices are labeled, and so Cayley's formula tells that the number of such trees is $n^{n-2}$. Solve the LP (Tutte \cite{Tutte}, Nash-Williams \cite{Nash-Williams}, Barahona \cite{Barahona}, Li et al. \cite{Li-et-al}):
\begin{eqnarray}
  \label{eqn:LP}
  \mbox{Maximize} & & \sum_{T \in \mathcal{T}_n} \lambda_T  \\
  \mbox{subject to} & (a) & \sum_{T \in \mathcal{T}_n : T \ni e} \lambda_T \leq C_e \quad \mbox{ for all } e \nonumber \\
  & (b) & \lambda_T \geq 0 \quad \quad \quad \quad \quad \mbox{ for all } T \in \mathcal{T}_n. \nonumber
\end{eqnarray}
Denote the maximum value of (\ref{eqn:LP}) as $\pi_n$. Then $\pi_n$ is the maximum rate at which node 1 can allcast its information to all the other nodes. The LP has a simple and intuitive explanation.
\begin{itemize}
\item  If one tags an infinitesimal information element originating at node 1 and follows the path of its spread to each of the $n-1$ recipients, one gets a directed graph rooted at the source node 1 and spanning all the $n$ nodes.
\item If the undirected version of this directed graph is not a tree, i.e., there is some cycle, then some node in the cycle is receiving this information element from two other nodes. One of these two incoming links can be removed without affecting the allcast property. We can thus reduce the directed graph to a {\em spanning arborescence}, which is a directed graph with no incoming links at the root node, exactly one incoming link at every other node, and all vertices are covered.
\item This spanning arborescence is in one-one correspondence with a tree, because the root is specified as node 1. So we may simply focus on the spanning tree associated with the arborescence. Call this tree $T$ (which is in $\mathcal{T}_n$).
\item Collect all information elements that are spread via this tree. Call its volume $\lambda_T$.
\end{itemize}
It is clear that each $\lambda_T \geq 0$ and constraint (a) in (\ref{eqn:LP}) is the capacity constraint associated with each of the links. Consequently, the value of the optimization problem in (\ref{eqn:LP}) is an upper bound on the optimal net flow from node 1. But it is immediate that any set of $\lambda_T$ satisfying the two constraints provides a means to achieve a rate $\sum_{T} \lambda_T$, since $\lambda_T$ units of information may be directed through the spanning arborescence associated with the tree $T$ and root vertex 1. Thus the maximum rate of allcast flow from a single sender is $\pi_n$, the solution to the LP in (\ref{eqn:LP}).

When link capacities are random, $\pi_n$ is a random variable whose asymptotics we shall soon characterize.

\subsection{Multicast}
\label{subsec:LP-multicast}

For the multicast problem, without loss of generality, let us index the session nodes as $\{ 1,2,\ldots,k_n \}$. As for allcast, by \cite[Cor.~4.a]{Li-et-al}, a multiple multicast rate vector $(r_1, r_2, \ldots, r_{k_n})$ with identical session nodes is achievable in an undirected capacitated network if and only if the rate vector $(\sum_{i=1}^{k_n} r_i, 0, \ldots, 0)$ is achievable, i.e., the sum rate is achievable for a {\em single} multicast with node 1 as sender and with the other $k_n-1$ nodes of the session as receivers\footnote{There is some subtlety involved here since, in general, network coding provides a coding advantage for multicasting in undirected networks; see \cite[Th. 4]{Li-et-al} for a proof of source independence in the single multicast case which is then generalized to get \cite[Cor. 4.a]{Li-et-al}}. We may therefore assume that there is but one sender, he is node 1, and all other $k_n-1$ nodes are recipients that must receive all information sent by node 1. Denote by $\mathcal{T}_n(k_n)$ the set of all Steiner trees that span the vertices $1,2,\ldots,k_n$. Obviously $\mathcal{T}_n(n) \equiv \mathcal{T}_n$. For multicast, again as for allcast, the maximum simultaneously transmissible rate from one sender (node 1) to the $k_n - 1$ other recipients is the maximum value of the modified LP (\cite{Li-et-al-2005}, \cite{Li-Li-Lau-2006}, \cite{Li-et-al}):
\begin{eqnarray}
  \label{eqn:LP-multicast}
  \mbox{Maximize} & & \sum_{T \in \mathcal{T}_n(k_n)} \lambda_T  \\
  \mbox{subject to} & (a) & \sum_{T \in \mathcal{T}_n(k_n) : T \ni e} \lambda_T \leq C_e \quad \mbox{ for all } e \nonumber \\
  & (b) & \lambda_T \geq 0 \quad \quad \quad \quad \quad \mbox{ for all } T \in \mathcal{T}_n(k_n). \nonumber
\end{eqnarray}
Set $\alpha_n = k_n / n$, and denote the maximum value of (\ref{eqn:LP-multicast}) as $\pi_n(\alpha_n)$. The above LP is the same as that of (\ref{eqn:LP}) with $\mathcal{T}_n$ replaced by the less restrictive $\mathcal{T}_n(k_n)$.

Again, when link capacities are random, $\pi_n(\alpha_n)$ is a random variable whose asymptotics we shall soon characterize.

\section{An Upper Bound}
\label{sec:upperbound}

Consider the following definitions.

\begin{itemize}
  \item Let $\chi_n$ and $\chi_n(k_n)$ denote the {\em maximum throughput achievable} in the allcast and multicast settings with the added possibility of network coding at each node. (The dependence of these quantities on the link capacities is understood and suppressed).
  \item Let $\eta_n$ denote the {\em strength} of the allcast network defined as follows. Let $\mathcal{P}$ denote the set of all partitions of the vertex set $\{ 1, 2, \ldots, n\}$. Consider a partition $\wp \in \mathcal{P}$. Let $\partial \wp$ denote the set of intercomponent links. Define
  \begin{equation}
    \label{eqn:eta-n-defn}
    \eta_n := \min_{\wp \in \mathcal{P}} \frac{\sum_{e \in \partial \wp} C_e}{|\wp| - 1}
  \end{equation}
  where $|\wp|$ denotes the number of subsets in the partition.
  \item Let $\eta_n(k_n)$ denote the strength of the multicast network with $k_n$ nodes in the session. This is defined as follows. Let $\mathcal{P}(k_n)$ denote the set of all partitions of the vertex set $\{1,2,\ldots,n\}$ such that each component of a partition contains at least one of the session nodes $\{1,2,\ldots,k_n\}$. Define
  \begin{equation}
    \label{eqn:eta-n-kn-defn}
    \eta_n(k_n) := \min_{\wp \in \mathcal{P}(k_n)} \frac{\sum_{e \in \partial \wp} C_e}{|\wp| - 1}.
  \end{equation}
\end{itemize}

Li et al. \cite{Li-et-al} showed the following result.


\begin{thm}{(Li et al. \cite[Th. 2 and Th. 3]{Li-et-al})} \\
(a) For any allcast session, $\pi_n = \chi_n = \eta_n$. \\
(b) For any multicast session, $\pi_n(k_n) \leq \chi_n(k_n) \leq \eta_n(k_n)$. $\hfill \IEEEQEDopen$
\label{thm:Li-et-al-relations}
\end{thm}

We can easily find good upper bounds on $\eta_n$ and $\eta_n(k_n)$ in random settings as shown in the following theorem.


\begin{thm}
 \label{thm:upperbound}
 Let $\{ C_{i,j} \}_{1 \leq i < j \leq n}$ denote the undirected link capacities. We then have the following upper bounds:
 \begin{eqnarray}
   \label{eqn:eta-n-bound}
   \eta_n & \leq & \frac{1}{n-1} \sum_{1 \leq i < j \leq n} C_{i,j} \\
   \label{eqn:eta-n-kn-bound}
   \eta_n(k_n) & \leq & \frac{1}{k_n-1}\left( \sum_{i < k_n} \sum_{j \geq k_n} C_{i,j} + \sum_{1 \leq i < j < k_n} C_{i,j} \right).
 \end{eqnarray}
 As a consequence, with $\lim_{n \rightarrow \infty} k_n/n = \alpha$, the inequalities
 \begin{eqnarray}
   \label{eqn:eta-n-limit-bound}
   \limsup_{n \rightarrow \infty} \frac{\eta_n}{n} & \leq & \frac{1}{2}\mathbb{E}[C] \\
   \label{eqn:eta-n-kn-limit-bound}
   \limsup_{n \rightarrow \infty} \frac{\eta_n(k_n)}{n} & \leq & \left( 1 - \frac{\alpha}{2} \right) \mathbb{E}[C]
 \end{eqnarray}
 hold almost surely. $\hfill \IEEEQEDopen$
\end{thm}


\begin{IEEEproof}
  Consider the partition $\wp = \{ \{1\}, \{2\}, \cdots, \{n\} \}$. There are $n$ subsets in the partition, and $\partial \wp$ is the set of all links. Apply now the definition (\ref{eqn:eta-n-defn}) of $\eta_n$ and we immediately get (\ref{eqn:eta-n-bound}) as the upper bound for the allcast case.

  For the multicast case, consider the partition
  \[
    \wp = \{ \{1\}, \{2\}, \cdots, \{k_n - 1\}, \{ k_n, \ldots, n \} \}.
  \]
  There are $k_n$ subsets in the partition. The set of links in $\partial \wp$ are
  \[
    \{ (i,j) : 1 \leq i < k_n, j \geq k_n \} \cup \{ (i,j) : 1 \leq i < j < k_n \}.
  \]
  Apply now the definition (\ref{eqn:eta-n-kn-defn}) of $\eta_n(k_n)$ and we immediately get (\ref{eqn:eta-n-kn-bound}) as the upper bound for the multicast case.

  Note that $|\partial \wp| = n(n-1)/2$ for allcast, and
  \begin{eqnarray}
    |\partial \wp| & = & (k_n - 1)(n-k_n+1) + \frac{ (k_n-1)(k_n-2) }{2} \nonumber \\
    \label{eqn:multicast-links}
    & = & (k_n-1) \left( n - \frac{k_n}{2} \right)
  \end{eqnarray}
  for multicast.

  Using $|\partial \wp| = n(n-1)/2$ for allcast in (\ref{eqn:eta-n-bound}), we obtain
  \[
    \frac{\eta_n}{n} \leq \frac{1}{2} \frac{1}{|\partial \wp|} \sum_{e \in \partial \wp} C_e.
  \]
  The sum on the right-hand side is composed of independent and identically distributed random variables. Consequently, the right-hand side converges almost surely to $\frac{1}{2} \mathbb{E}[C]$ by the strong law of large numbers, and we obtain (\ref{eqn:eta-n-limit-bound}).

  For the multicast case, use (\ref{eqn:multicast-links}) in (\ref{eqn:eta-n-kn-bound}) to obtain
  \[
    \frac{\eta_n(k_n)}{n} \leq \left( 1 - \frac{k_n}{2n} \right) \frac{1}{|\partial \wp|} \sum_{e \in \partial \wp} C_e.
  \]
  Again by an application of the strong law of large numbers, the conclusion (\ref{eqn:eta-n-kn-limit-bound}) follows.
\end{IEEEproof}


Observe that, by Theorem \ref{thm:Li-et-al-relations}, the upper bounds in Theorem \ref{thm:upperbound} apply for capacity with the possibility of network coding. Let us now turn to achievability of these rates in their respective settings.

\section{Allcast: Achievability}
\label{sec:allcast-achievability}

In this section we consider the allcast setting and argue that the upper bound in (\ref{eqn:eta-n-limit-bound}) is tight, and moreover, the upper bound is achievable via flows. After first establishing the existence of a scheme, we then provide a practical decentralized asymptotically optimal push-pull algorithm.


\begin{thm}
  \label{thm:allcast}
  For the allcast problem, we have
  \[
    \lim_{n \rightarrow \infty} \frac{\pi_n}{n} = \frac{1}{2} \mathbb{E}[C] \quad \mbox{ a.s.}
  \]
  $\hfill \IEEEQEDopen$
\end{thm}


\begin{IEEEproof}
The fact that we cannot do better than $\mathbb{E}[C]/2$ was already established in (\ref{eqn:eta-n-limit-bound}). So the proof of the above theorem would be complete if we can establish that $\mathbb{E}[C]/2$ is achievable. We first argue achievability on the simpler Erd\H{o}s-R\'{e}nyi graphs. We then lift this result to the general case.

Take the random graph $G(n,p)$ where each link capacity is iid with Bernoulli($p$) distribution. Catlin et al. \cite[Sec. 3]{Catlin-et-al} proved the stronger result that, even if $p$ vanishes with $n$, so long as it is larger than $(28 \log n / n)^{1/3}$, we have for all sufficiently large $n$ the equality
\begin{equation}
  \label{eqn:Catlin-actual}
  \pi_n = \left\lfloor \frac{\sum_{1 \leq i < j \leq n} C_{i,j}}{n-1} \right\rfloor \quad \mbox{ a.s.}
\end{equation}
For any $\varepsilon > 0$, using $p > 0$, the result in (\ref{eqn:Catlin-actual}), and the strong law of large numbers, we have
\begin{equation}
  \label{eqn:Catlin}
  \liminf_{n \rightarrow \infty} \frac{\pi_n}{n} \geq \frac{p}{2} (1 - \varepsilon) \quad \mbox{ a.s.}
\end{equation}
By excluding all null sets associated with rational $\varepsilon \in (0,1)$, it follows that
\[
  \liminf_{n \rightarrow \infty} \frac{\pi_n}{n} \geq \frac{p}{2} \quad \mbox{ a.s.}
\]

There now remains the step of lifting this result to any generic distribution $F$, for the iid capacities $C_{i,j}$, satisfying
\begin{equation}
  \label{eqn:finite-expectation}
  0 < \mathbb{E}[C] = \int_0^{\infty} \Pr \{ C > x \} ~dx = \int_0^{\infty} [1 - F(x)] ~dx < \infty.
\end{equation}
This is readily done. Fix an arbitrary $\varepsilon > 0$. By (\ref{eqn:finite-expectation}) and the fact that the function $1 - F(x)$ is Riemann integrable (for it is Lebesgue integrable, bounded, and has at most a countable number of discontinuities), we can choose a natural number $M < \infty$ and  $\delta > 0$ such that
\begin{eqnarray}
  \label{eqn:expectation-approximation}
  \sum_{k=1}^M \delta \cdot [1 - F( k \delta) ] & \geq & \mathbb{E}[C] \cdot (1 - \varepsilon).
\end{eqnarray}
We now build a family of $M$ coupled graphs, each with $n$ vertices. For a realization of the iid link capacities, let $G_k$ be a new graph on the $n$ vertices with link between $i$ and $j$ if and only if $C_{i,j} > k \delta$, for $k = 1, 2, \ldots, M$. Clearly, $G_k$ is an Erd\H{o}s-R\'{e}nyi graph on $n$ vertices with parameter
\[
  p(k) := \Pr \{ C > k\delta \} = 1 - F( k \delta ).
\]
On $G_k$, we interpret each link, if present, as having capacity $\delta$. While the graphs are coupled across the parameter $k$, for a fixed $k$, the links on the graph $G_k$ are iid Bernoulli($p(k)$) random variables. Let $\pi_n(G_k)$ be the maximum number of disjoint trees that can be packed in $G_k$. By the result (\ref{eqn:Catlin}) applied to each fixed $k$, we have
\begin{eqnarray*}
  \liminf_{n \rightarrow \infty} \frac{\pi_n}{n} & \geq & \liminf_{n \rightarrow \infty} \frac{1}{n} \sum_{k=1}^M \delta \cdot \pi_n(G_k) \\
  & \geq & \delta \cdot \sum_{k=1}^M \frac{p(k)}{2} (1 - \varepsilon ), \quad \mbox{ a.s.} \\
  & = & \frac{1}{2} \sum_{k=1}^M \delta \cdot [1 - F ( k \delta)] \cdot (1 - \varepsilon) \\
  & \geq & \frac{1}{2} \cdot \mathbb{E}[C] \cdot (1 - \varepsilon) \cdot (1 - \varepsilon)  \\
  & \geq & \frac{\mathbb{E}[C]}{2}(1 - 2\varepsilon),
\end{eqnarray*}
where the penultimate inequality follows from (\ref{eqn:expectation-approximation}). It follows as before that $\lim_n \frac{\pi_n}{n} \geq \frac{\mathbb{E}[C]}{2}$ almost surely. This completes the proof. (See \cite{200909ORL_AldMcdSco} or \cite{Khandwawala-Sundaresan} for a similar truncation, quantization, and scaling argument).
\end{IEEEproof}


The key to proving Theorem \ref{thm:allcast} is the result (\ref{eqn:Catlin}) on Erd\H{o}s-R\'{e}nyi graphs. In order to show this, we utilized the result (\ref{eqn:Catlin-actual}) of Catlin et al. \cite{Catlin-et-al}. The main point of the rest of this section is to demonstrate that (\ref{eqn:Catlin}) can be proved constructively using a rather simple and decentralized algorithm.

\subsection{{\em \textsf{\small ALLCAST}}: A decentralized algorithm for allcast in a random graph}
\label{subsec:allcast-allgorithm}

\begin{figure}[bt]
    \centering
    \includegraphics[scale=0.35]{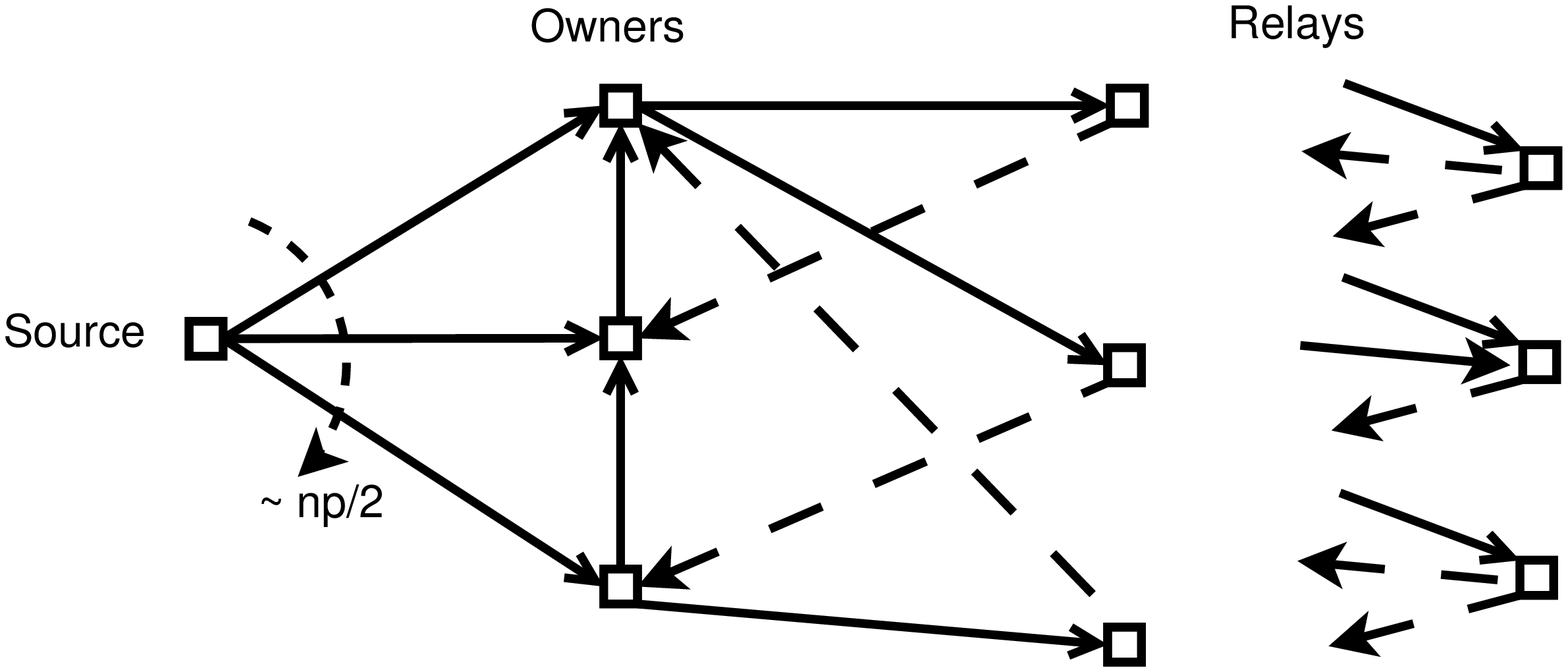}
    \caption{Graph showing the three sets of nodes: source, owners, and relays. Source pushes bits to owners who then push to relays. All nodes then pull from owners and any remaining bits from relays.}
    \label{fig:allcast-algorithm}
\end{figure}

This section describes a decentralized {\em push-pull} algorithm for allcast that achieves (\ref{eqn:Catlin}) for an arbitrary $\varepsilon > 0$. For ease of exposition, we shall assume a total of $n+1$ nodes with node 0 as the source node. The source node 0 has to push a total of $\frac{1}{2}np(1 - \varepsilon)$ bits to all nodes. We have ignored integer rounding and a factor $(n+1)/n$ both of which are easily absorbed into $\varepsilon$. The algorithm broadly has two push steps and two pull steps, as described next. See Figure \ref{fig:allcast-algorithm}. The analysis that comes later will argue that with overwhelming probability none of the steps fail. \\

\begin{itemize}[{\bf Algorithm} \textsf{\small ALLCAST}:]
  \item {\em Setting up of directions}: All links that do not involve the source node 0 are assigned one of the two directions with equal probability, independently of the choices of directions at other links. All links that involve the source node 0 have a direction pointing away from the source.
  \item {\em Push step 1}: Source node 0 {\em pushes} $\frac{1}{2}np(1 - \varepsilon)$ different bits to that many of its neighbors. We number the bits $b_1, b_2, \ldots, b_{np(1-\varepsilon)/2}$, call the respective recipient nodes as {\em owners} of these bits, and denote the owners (sometimes) as $O_1, O_2, \ldots, O_{np(1-\varepsilon)/2}$ instead of saying node 1, node 2, \ldots, node $np(1-\varepsilon)/2$. There may be several other neighbors of node 0, but the corresponding links are left unused. These and other nodes who are not owners are called {\em relays}, and are denoted $R_{np(1-\varepsilon)/2 + 1}, \ldots, R_n$ (instead of saying node $np(1-\varepsilon)/2 + 1$, \ldots, node $n$).
  \item {\em Push step 2}: Each owner $O_i$ pushes his bit $b_i$ one more level along links that point outward from $i$, regardless of the status of the recipient as an owner of another bit or a relay. The receiving node will then have $b_i$ (and similarly many other bits) for other nodes to pull in the next couple of steps of the algorithm.
  \item {\em Pull step 1}: Each node, say node $j$, collects all incoming bits $b_i$ coming directly from owners $O_i$ via links $i \rightarrow j$. (This is the bit pushed by $O_i$ in push step 2).
  \item {\em Pull step 2}: Having collected some bits directly from owners, node $j$ identifies the remaining bits, the relays to which it is connected with direction pointing towards $j$, and the bits that these relays have available having received the bits directly from owners. A representation of this information is the {\em bit-map} matrix of nodes and bits they have available for pulling (see Table \ref{tb:allcast-bitmap} and its description). Node $j$ then identifies a {\em complete matching} of these desired bits to the helper relays: each desired yet-to-be-pulled bit is pulled from a suitable relay that has the bit, with each relay accounting for one bit, and this constitutes a matching. $\hfill \IEEEQEDopen$
\end{itemize}

\begin{table*}[t]
\begin{center}
\caption{Allcast bit-map}
\label{tb:allcast-bitmap}
{\small
\hfill{}
\begin{tabular}{|c|cccccccccc||ccccccc|} \hline \hline
        &&&&&&&&&&&&&&&&& \\
        &$O_1$  &$O_2$ & $\cdots$ &$O_t$ &$\cdots$ &$O_a$ &$O_b$ &$O_c$ &$\cdots$ &$O_{\frac{np(1-\varepsilon)}{2}}$ &$R_{\frac{np(1-\varepsilon)}{2}+1}$ &$\cdots$ &$R_u$ &$R_v$ &$R_w$ &$\cdots$ &$R_n$ \\
        &$b_1$  &$b_2$ & $\cdots$ &$b_t$ &$\cdots$ &$b_a$ &$b_b$ &$b_c$ &$\cdots$ &$b_{\frac{np(1-\varepsilon)}{2}}$ &&&&&&& \\
        &&&&&&&&&&&&&&&&& \\ \hline \hline
        &&&&&&&&&&&&&&&&& \\
 $O_1$  &1  & &$\cdots$  &  &   &$X_{1a}$ & & &$\cdots$ & & & & & & & &  \\
 $O_2$ & &1 & & & & & & & & & & & & & & & \\
 $\vdots$ & & &$\ddots$ & & & & & & & & & & & & & & \\
 $O_t$  &1 &1 &$\cdots$ &1 &$\cdots$ &0 &0 &0 &$\cdots$ &1 &0 &$\cdots$ &1 &1 &1 &$\cdots$ &0 \\
 $\vdots$ & & & & &$\ddots$ & & & & & & & & & & & & \\
 $O_a$  &$X_{a1}$ & &$\cdots$ & & &1 & & &$\cdots$ & &$X_{ai}$ & & & & & &  \\
 $O_b$  & & & & & & &1 & & & & & & & & & & \\
 $O_c$  & & & & & & & &1 & & & & & & & & & \\
 $\vdots$ & & & & & & & & &$\ddots$ & & & & & & & & \\
 $O_{\frac{np(1-\varepsilon)}{2}}$  & & & & & & & & & &1 & & & & & & & \\
 & & & & & & & & & & & & & & & & & \\
 \hline \hline
 & & & & & & & & & & & & & & & & & \\
 $R_{\frac{np(1-\varepsilon)}{2}+1}$   & & & & & &$X_{ia}$ & & & & & & & & & & & \\
 $\vdots$ & & & &$\vdots$ & & & & & & & & & & & & & \\
 $R_u$  & & & &0 &$\cdots$ & &1 & &$\cdots$ & & & & & & & & \\
 $R_v$  & & & &0 &$\cdots$ & & &1 &$\cdots$ & & & & & & & & \\
 $R_w$  & & & &0 &$\cdots$ &1 & & &$\cdots$ & & & & & & & & \\
 $\vdots$ & & & &$\vdots$ & & & & & & & & & & & & & \\
 $R_n$  & & & & & & & & & &1 & & & & & & & \\
         &&&&&&&&&&&&&&&&& \\
 \hline
 \hline
\end{tabular}
}
\end{center}
\end{table*}


The orientation step (the first step of the algorithm), when operating on a node other than the source, renders roughly one half of the links outward and the remaining links inward. The outward links provide service to other nodes. The inward links bring in the $np(1-\varepsilon)/2$ bits to the node. In this sense, the resource usages for rendering service and reaping benefit are balanced.

Before we dive into an analysis of this algorithm, we describe the bit-map of Table \ref{tb:allcast-bitmap} in more detail. The rows and columns are indexed as
\[
  O_1, O_2, \ldots, O_{np(1-\varepsilon)/2}, R_{np(1-\varepsilon)/2}+1, \ldots, R_n.
\]
In addition, the first $np(1-\varepsilon)/2$ columns will also refer to the corresponding bits.
\begin{itemize}
  \item For $1 \leq i \leq np(1-\varepsilon)/2$, we write $X_{i,i} = 1$ to signify that node $O_i$ has bit $b_i$.
  \item For $i \neq j$, since the link $\{i,j\}$ itself occurs with probability $p$, and further, may have either direction with equal probability, we have
  \begin{eqnarray*}
    X_{i,j} = 1, X_{j,i} = 0 & & \mbox{ if } j \rightarrow i; \\
    X_{j,i} = 0, X_{i,j} = 1 & & \mbox{ if } i \rightarrow j; \\
    X_{j,i} = 0, X_{i,j} = 0 & & \mbox{ if no link between $i$ and $j$}.
  \end{eqnarray*}
  These are mutually exclusive, with the first setting occurring with probability $p/2$, the second setting with probability $p/2$, and the third setting with probability $1-p$.
  \item If $X_{i,j} = 1$, then node $i$ (owner or relay) can obtain bit $b_j$ from owner $O_j$ (if $1 \leq j \leq np(1-\varepsilon)/2$) or some bit that relay $R_j$ has (if $j > np(1-\varepsilon)/2$).
  \item The set of bits node $i$ receives directly from owners corresponds to the set of 1s in the first $np(1-\varepsilon)/2$ columns of the $i$th row, for if $X_{i,j} = 1$, then owner $O_j$ pushes his bit $b_j$ to node $i$. (For example, in Table \ref{tb:allcast-bitmap}, owner $O_t$ has bits $b_1, b_2, b_t, b_{np(1-\varepsilon)/2}$, but does not have $b_a, b_b, b_c$).
  \item The 1s in the $i$th row beyond column $np(1-\varepsilon)/2$ point to relays that can be used by node $i$ to pull any remaining bits in pull step 2. (For example, owner $O_t$ is connected to relays $R_u, R_v, R_w$ with directions pointing towards $O_t$. These relays will help node $O_t$ get the yet-to-be-pulled bits $b_a, b_b, b_c$).
  \item Clearly, while the random variables $X_{i,j}$ and $X_{j,i}$ are coupled, the nondiagonal entries of the $i$th row
  \[
    \{ X_{i,j}, 1 \leq j \leq n, j \neq i \}
  \]
  are iid Bernoulli($p/2$) random variables, for $1 \leq i \leq n$. The same holds for nondiagonal entries of any column.
\end{itemize}

Our main assertion is that the algorithm \textsf{\small ALLCAST} succeeds with high probability in distributing the $np(1-\varepsilon)/2$ bits to all nodes.


\begin{thm}
\label{thm:allcast-algo-asym}
For any $\varepsilon > 0$, the following event occurs almost surely: for all but finitely many $n$, the algorithm \textsf{\small ALLCAST} succeeds in distributing all $np(1-\varepsilon)/2$ bits to each of the $n$ nodes. $\hfill \IEEEQEDopen$
\end{thm}


{\em Remarks}: 1) It follows immediately that, for any $\varepsilon > 0$, the inequality (\ref{eqn:Catlin}) holds.

2) The above theorem also implies that, for all sufficiently large $n$, we can pack $np(1-\varepsilon)/2$ disjoint (spanning) trees in $G(n,p)$, with each tree having the property that it has depth at most 3.

3) \textsf{\small ALLCAST} is decentralized in the following sense. The direction of each link, when present and if the source node is not involved, is picked at random by the toss of a fair coin, and this information is needed only at these two incident nodes. The two levels of pushes, and thus the first pull stage, are easily seen to be decentralized. At each node, the actions depend only on the links incident on it and the agreed upon link directions. Each node then keeps a list of bits it receives from owners. For the final pull stage, each node has to get this list associated with each of its potential helper relays. This is the step that may involve significant exchange of information, but the cost involved is a one-time set-up cost that can be amortized over multiple rounds of data communication. Note that all information exchanges (link directions, pushing of owned bits, lists of bits available at neighboring helper relays) are of information which are of local relevance that are, in addition, locally available. The matching can be identified in $O(n^2)$ steps \cite{Karp-et-al}.

4) We need three elementary tools to establish the result. The first is the following well known concentration result for the binomial distribution, which we state without proof.


\begin{lemma}
  \label{lem:binomial-concentration}(\cite[Th. 1.7(i)]{Bollobas})
  Suppose $0 < q < \frac{1}{2}$, $0 < \varepsilon < 1/12$, and $\varepsilon n q (1-q) \geq 12$. Let $S_{n,q}$ be the sum of $n$ Bernoulli($q$) random variables. Then
  \begin{equation}
    \Pr \left\{ \left| \frac{1}{nq} S_{n,q} - 1 \right| > \varepsilon \right\} \leq \frac{1}{\sqrt{\varepsilon^2 nq}} e^{-nq\varepsilon^2/3}.
  \end{equation}
  $\hfill \IEEEQEDopen$
\end{lemma}


This result holds for every $n$ and $q$ satisfying $\varepsilon n q (1-q) \geq 12$, and as such, $q$ can vary with $n$. The second tool is the Borel-Cantelli lemma that gives us a sufficient condition for almost sure convergence. The third tool is one of existence of matchings on random bipartite graphs, which will be the subject of Section \ref{sec:isolated-node-no-matching}.

\begin{IEEEproof}[Proof of Theorem \ref{thm:allcast-algo-asym}]
By the Borel-Cantelli lemma, it suffices to show that the probability that the algorithm fails for a particular $n$ is summable over $n$. If the algorithm fails, then at least one of the following is true.

1) The event $A_1^{(n)}$ occurs, which is defined to be the event that there are fewer than $\frac{1}{2}np(1-\varepsilon)$ vertices connected to node 0. By Lemma \ref{lem:binomial-concentration}, there is some $c_1 > 0$ such that for all sufficiently large $n$, we have $\Pr\{ A^{(n)}_1 \} \leq e^{-c_1 n}$.

2) For some node $t$, the event $A^{(n)}_2(t)$ occurs, which is defined to be the event that the node $t$ is connected to a certain number of owners outside the range $\frac{1}{2}np(1-\varepsilon) \cdot \frac{1}{2} p (1 \pm \varepsilon)$ with links pointing towards $t$. (If node $t$ is an owner, there are $\frac{1}{2}np(1-\varepsilon)) - 1$ other owners, but the 1 can be absorbed into the $(1-\varepsilon)$ factor). Again by Lemma \ref{lem:binomial-concentration}, there is some $c_2 > 0$ such that for all sufficiently large $n$, we have $\Pr\{ A^{(n)}_2(t) \} \leq e^{-c_2 n}$.

3) For some node $t$, the event $A^{(n)}_3(t)$ occurs, which is the event that the node $t$ is connected to fewer than
  \begin{eqnarray*}
    \beta_n & := & \left( n - \frac{1}{2}np(1-\varepsilon) \right) \cdot \frac{1}{2} p (1-\varepsilon) \\
            & = & \frac{1}{2}np(1-\varepsilon) \cdot \left( 1- \frac{1}{2}p(1-\varepsilon) \right)
  \end{eqnarray*}
  relays with links pointing towards $t$. (Again, the case of 1 less relay when node $t$ is a relay is easily handled). Once again by Lemma \ref{lem:binomial-concentration}, there is a $c_3 > 0$ such that for all sufficiently large $n$, we have $\Pr\{ A^{(n)}_3(t) \} \leq e^{-c_3 n}$.

4) For some node $t$, if $A_1^{(n)} \cup A^{(n)}_2(t) \cup A^{(n)}_3(t)$ does not occur, then the event $M^{(n)}(t)$ occurs, which is the event that node $t$ is unable to pull the desired bits. We claim that
      \begin{equation}
        \label{eqn:no-matching}
        \Pr \left\{ M^{(n)}(t) ~|~ \left( A_1^{(n)} \cup A^{(n)}_2(t) \cup A^{(n)}_3(t) \right)^c \right\} \leq \gamma(\beta_n)
      \end{equation}
      for some sequence $\gamma : \mathbb{N} \rightarrow [0,1]$ satisfying
      \begin{equation}
        \label{eqn:no-matching-summable}
        \sum_{n=1}^{\infty} n \gamma(\beta_n) < \infty.
      \end{equation}

The event that the algorithm fails is then a subset of
\[
  A_1^{(n)} \bigcup_{t=1}^n \left( A^{(n)}_2(t) \cup A^{(n)}_3(t) \cup M^{(n)}(t)\right)
\]
whose probability is upper bounded via the union bound and (\ref{eqn:no-matching}) by
\[
   n \cdot \left( e^{-nc_1} + e^{-nc_2} + e^{-n c_3} + \gamma(\beta_n) \right)
\]
which, by the summability claim in (\ref{eqn:no-matching-summable}) and the exponentially decaying nature of the other terms, is summable.

Let us now prove (\ref{eqn:no-matching}) and (\ref{eqn:no-matching-summable}).

Fix a node $t$, where $1 \leq t \leq n$. The event $A^{(n)}_1$ has not occurred, and so the source has sent out exactly $\frac{1}{2}np(1-\varepsilon)$ bits to that many owners. The event $A^{(n)}_2(t)$ has not occurred, and so node $t$ is connected to between $\frac{1}{2}np(1-\varepsilon) \cdot \frac{1}{2}p(1 \pm \varepsilon)$ owners with links towards node $t$. The connected owners directly furnish their bits to node $t$. But node $t$ needs at least $\frac{1}{2}np(1-\varepsilon) - \frac{1}{2}np(1 - \varepsilon) \cdot \frac{1}{2}p(1 + \varepsilon)$ additional bits to be pulled in pull step 2. This set of yet-to-be-pulled bits points to some random selection of columns from amongst the first $\frac{1}{2}np(1-\varepsilon)$ columns and does not include column $t$.

The event $A^{(n)}_3(t)$ has not occurred, and so node $t$ is connected to at least $\beta_n$ relays that could potentially furnish these missing bits (that is, with links towards node $t$). Consider the rows corresponding to these relays. This set of rows is a random selection of at least $\beta_n$ rows from amongst the indices $\frac{1}{2}np(1-\varepsilon) + 1$ through $n$ and does not include $t$.

Observe that conditioned on these selections, the entries of the submatrix continue to be iid Bernoulli($p/2$) random variables. If $M^{(n)}(t)$ occurs, there is no coverage of these the yet-to-be-pulled bits (columns) using the helper relays (rows), with each helper relay furnishing at most one missing bit. But this in particular implies that there is no coverage of the yet-to-be-pulled bits (columns) by some subset of exactly $\beta_n$ helper relays (rows) with each helper relay furnishing at most one bit. But this further implies that any superset of $\beta_n$ columns that includes the yet-to-be-pulled bits (columns), and continues to exclude column $t$, cannot be {\em matched} to the selected $\beta_n$ helper relays (rows). Now, Lemma \ref{lem:summability} of Section \ref{sec:isolated-node-no-matching} shows that this probability is upper bounded by $\gamma(\beta_n)$, which is (\ref{eqn:no-matching}), and that $n \gamma(\beta_n)$ is summable, which is (\ref{eqn:no-matching-summable}). This concludes the proof.
\end{IEEEproof}


The matching step above is the key to complete the deliveries. It ensures  that all required bits are available at some helper relay, and that each link has at most 1 bit load so that capacity constraints are not violated. We now devote a section to demonstrating this key step.

\section{The existence of a bipartite matching}
\label{sec:isolated-node-no-matching}

In this section, we establish the crucial step of existence of bipartite matchings. The following lemma, taken from Bollob\'{a}s \cite{Bollobas}, is key to showing that matchings exist almost surely and one can pull the $\beta_n$ bits from relays. We first present the result for a random bipartite graph with $n$ vertices on each side. The results of this section are well-known and are provided only for completeness and ease of reference.


\begin{lemma}(\cite[Lem. 7.12, p. 174]{Bollobas}).
\label{lem:Bollobas-consequence}
Let $G$ be a bipartite graph with vertex sets $V_1, V_2$ such that $|V_1| = |V_2| = n$. Suppose $G$ does not have any isolated vertices and it does not have a complete matching. Then there is a set $A \subset V_i$ for either $i = 1$ or $2$ such that the following three conditions hold:
    \begin{itemize}
      \item[(i)] $\Gamma(A)$ has $|A|-1$ elements,
      \item[(ii)] the subgraph spanned by $A \cup \Gamma(A)$ is connected,
      \item[(iii)] $2 \leq |A| \leq (n+1)/2$. $\hfill \IEEEQEDopen$
    \end{itemize}
\end{lemma}


The above conditions are simple consequences of Hall's marriage theorem and some elementary observations. The proof can be found in \cite[Lem. 7.12, p. 174]{Bollobas}. We now bound the probability of these events on a random bipartite graph $G(n,n,p)$ (see Section \ref{subsec:randomness-model}).


\begin{lemma}
 \label{lem:Bollobas-consequence2}
 Let $F_a$ be the event that there is a set $A$ of size $a$ with $A \subset V_i$ for $i = 1$ or 2 satisfying (i)-(iii) of Lemma \ref{lem:Bollobas-consequence}. Let $n_1 = (n+1)/2$. Consider $G(n,n,p)$. Then $\Pr \{ \cup_{a=2}^{n_1} F_a \} \leq \varepsilon_n$ where $\varepsilon_n$ summable, and hence $\varepsilon_n \rightarrow 0$. Furthermore, we also have $\sum_{n \geq 1} n \varepsilon_n < \infty$. $\hfill \IEEEQEDopen$
\end{lemma}


\begin{IEEEproof}
 Fix $a$. There are two choices for $i$ in the condition $A \subset V_i$, there are $\binom{n}{a}$ ways to choose the subset $A$, and there are $\binom{n}{a-1}$ ways to choose the subset $\Gamma(A)$. Once chosen, there must be no links between the $a$ vertices of $A$ and the $n-a+1$ vertices of $V_2 - \Gamma(A)$. By the union bound (for the possibilities for $A$ and $\Gamma(A)$), we get
  \begin{equation}
    \label{eqn:Fa-bound}
    \Pr \{ F_a \} \leq 2 \binom{n}{a} \binom{n}{a-1} (1-p)^{a(n-a+1)}.
  \end{equation}
  Using $\binom{n}{a} \leq n^a$, by a second application of the union bound, and by dropping some factors that are smaller than 1, we get
  \begin{equation}
    \label{eqn:epsilon_n-defn}
    \Pr \{ \cup_{a=2}^{n_1}F_a \} \leq 2 \sum_{a=2}^{n_1} n^{2a-1} (1-p)^{an} (1-p)^{-a^2} =: \varepsilon_n.
  \end{equation}
  For an $a_0$, set $n_0 = 2a_0 - 1$. It suffices to show that for $n_0$ large, $\sum_{n \geq n_0} \varepsilon_n < \infty$. Interchanging the indices of summation, and changing limits appropriately, we get
  \begin{eqnarray}
    \sum_{n \geq n_0} \varepsilon_n & = & 2\sum_{a=2}^{a_0} (1-p)^{-a^2} \sum_{n \geq n_0} n^{2a-1} (1-p)^{an} \nonumber\\
     && + ~2\sum_{a > a_0} (1-p)^{-a^2} \sum_{n \geq 2a - 1} n^{2a-1} (1-p)^{an}. \nonumber \\
    \label{eqn:epsilon_n-sum}
    &&
  \end{eqnarray}
  The first term is easily seen to be summable for any finite $a_0$. For the second one, observe that for any $\delta > 0$ and any $C > 0$, there is an $a_0$ large enough so that for all $a > a_0$ and all $n \geq 2a-1$, we have $n^{2a-1} \leq n^{2a} \leq C (1+\delta)^{an}$. By taking $C = (1-p)(1-\delta) (1 - (1-p)(1-\delta))$ it follows that
  \[
    \sum_{n \geq 2a - 1} n^{2a-1} (1-p)^{an} \leq (1-p)^{2a^2}(1+\delta)^{2a^2}.
  \]
  Choose $\delta$ small enough so that $(1-p) (1+\delta)^2 < 1$. Substitute this in the second term in (\ref{eqn:epsilon_n-sum}), and we see that it is summable.

  Finally, to show that $\sum_{n \geq 1} n \varepsilon_n < \infty$, we modify (\ref{eqn:epsilon_n-sum}) as
  \begin{eqnarray*}
    \sum_{n \geq n_0} n \varepsilon_n & = & 2\sum_{a=2}^{a_0} (1-p)^{-a^2} \sum_{n \geq n_0} n^{2a} (1-p)^{an} \\
     & & + ~ 2\sum_{a > a_0} (1-p)^{-a^2} \sum_{n \geq 2a - 1} n^{2a} (1-p)^{an}.
  \end{eqnarray*}
  By our choice of $a_0$ and $\delta$, we also have $n^{2a} \leq C(1+\delta)^{an}$, and so all the steps that followed (\ref{eqn:epsilon_n-sum}) apply, which establishes summability of $n\varepsilon_n$.
\end{IEEEproof}


We now put these together to argue that a bipartite matching exists in $G(n,n,p)$ with high probability.


\begin{thm}
\label{thm:bipartite-matching-n}
The probability that $G(n,n,p)$ does not have a complete matching is upper bounded by $\gamma(n) := 2n(1-p)^n + \varepsilon_n$, where $\varepsilon_n$, defined in (\ref{eqn:epsilon_n-defn}), has all the properties indicated in Lemma \ref{lem:Bollobas-consequence2}. $\hfill \IEEEQEDopen$
\end{thm}


\begin{IEEEproof}
If $G(n,n,p)$ does not have a complete matching, then either (1) there is an isolated vertex, or (2) there is no isolated vertex and by virtue of Lemma \ref{lem:Bollobas-consequence}, $\cup_{a=1}^{n_1} F_a$ must occur, where $n_1 = (n+1)/2$ as before. By Lemma \ref{lem:Bollobas-consequence2}, the probability of the second case event is at most $\varepsilon_n$. The probability that there is no isolated vertex is, by the union bound, at most $2n(1-p)^n$.
\end{IEEEproof}


In the previous section, we had a need to study existence of bipartite matchings over left and right sets of size $\beta_n := \lfloor cn \rfloor$ where $0 < c < 1$.


\begin{lemma}
\label{lem:summability}
For a fixed $0 < c < 1$, let $\beta_n := \lfloor cn \rfloor$. The probability that $G(\beta_n, \beta_n, p)$ does not have a complete matching is upper bounded by $\gamma(\beta_n)$ where $\gamma$ is the upper bounding function defined in Theorem \ref{thm:bipartite-matching-n}. Furthermore, $\sum_{n \geq 1} n \gamma(\beta_n) < \infty$. $\hfill \IEEEQEDopen$
\end{lemma}


\begin{IEEEproof}
The upper bound on the probability that a matching does not exist is immediate. We now show that $\sum_n n \gamma(\beta_n)$ converges. Note that any particular integer repeats at most $1/c+1$ times in the sequence $\{ \beta_n, n \geq 1 \}$. As a consequence
\begin{eqnarray*}
  \sum_{n \geq 1} n \gamma(\beta_n) & \leq & \frac{1}{c} \sum_{n \geq 1} (cn) \cdot \gamma(\beta_n) \\
    & \leq & \frac{1}{c} \sum_{n \geq 1} (\beta_n + 1) \cdot \gamma(\beta_n) \\
    & \leq & \frac{1}{c} \left( \frac{1}{c} + 1 \right) \sum_{k \geq 1} (k + 1) \cdot \gamma(k)  ~ < ~ \infty.
\end{eqnarray*}
\end{IEEEproof}

\section{A Digression of Not Just Interpretive Value: Maximum Single Commodity Flow}
\label{sec:single-commodity}

Let us now take a step back to see how matching arises naturally in the simpler case of a single commodity flow between a source node $s$ and a sink node $t$. We shall assume that additional nodes $1, 2, \ldots, n$ are merely relays. The random graph of interest is now $G(n+2, p)$, where the number $n+2$ comes from $n$ relay nodes and the two source and sink nodes. Our interest is in the maximum rate of information flow between source and sink $\pi_n(2)$. (To be strictly conforming to our earlier notation, we must use $\pi_{n+2}(2)$ for there are $n+2$ nodes in the network and with the first two nodes being in session. The asymptotics does not change of course).

\begin{figure}[tb]
\centering
\includegraphics[width=3.4in, height=1.7in]{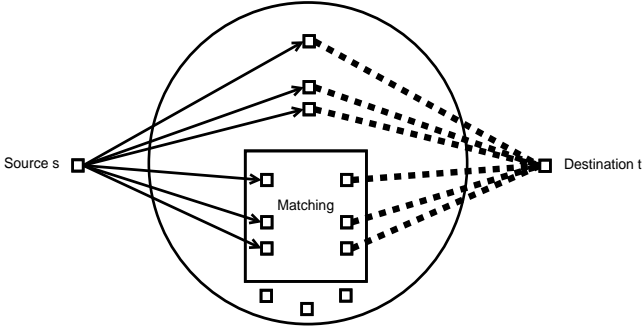}
\caption{Single source single sink setting indicating how matching arises.}
\label{fig:maxflow-kmn}
\end{figure}

Grimmett and Suen \cite{Grimmett-Suen} showed that $\pi_n(2)$ grows linearly in $n$ and that $\lim_{n} \frac{\pi_n(2)}{n} = p$, almost surely. It is then clear that the cut that isolates the source is a tight cut. So is the cut that isolates the sink. Motivated by this, Karp et al. \cite{Karp-et-al} provided an algorithm that achieves the minimum cut capacity. We will show that, for a fixed $\varepsilon > 0$, the following algorithm transports $np(1-\varepsilon)$ bits from the source to the sink with vanishing probability of failure. See Figure \ref{fig:maxflow-kmn}.

\vspace*{.1in}

\begin{itemize}[{\bf Algorithm} \textsf{\small MaxFlow}:]
  \item The source floods exactly $np(1-\varepsilon)$ links with one bit per link.
  \item The sink pulls all these bits from $np(1-\varepsilon)$ links connected to it in the following two steps.

  ~(a) If any node connected to the sink is directly connected to the source, the sink draws the corresponding bit. With overwhelming probability, there are at least $np(1-\varepsilon) \cdot p(1-\varepsilon)$ such connections.

  ~(b) Here is how the sink draws the remaining bits. There are at most $\beta_n = np(1-\varepsilon)(1 - p(1-\varepsilon))$ such yet-to-be-pulled bits, and these reside with let us say {\em source side relays} not in direct contact with the sink. Among those relays that did not get a bit directly from the source (and these are $n - np(1-\varepsilon) = n(1-p(1-\varepsilon))$ in number) the sink is connected to at least $n(1-p(1-\varepsilon)) \cdot p(1-\varepsilon) = \beta_n$, again with overwhelming probability. Let us call these the {\em sink side relays}. There is a matching, again with overwhelming probability, between the source side relays and the sink side relays. This matching is then used in the obvious way to draw the yet-to-be-pulled bits. $\hfill \IEEEQEDopen$
\end{itemize}


Obviously, the direct link between $s$ and $t$ is inconsequential for the asymptotics. It is further obvious from the analysis of the previous section that the probability of failure is overwhelmingly small, and moreover, it is summable over $n$ (Lemma \ref{lem:summability}). This is essentially the argument of Karp et al. \cite{Karp-et-al} to show the achievability direction of the result of Grimmett and Suen \cite{Grimmett-Suen}.

What if we have not one sink $t$, but two sinks $t_1$ and $t_2$? There is one matching needed for $t_1$ and another needed for $t_2$. These matchings depend on the connections at the respective sinks, but can be found with overwhelmingly small probability of failure via the union bound for probabilities. Once these are found, while the relays may be overworked, the links are utilized within their capacity limits. Indeed, if a common sink-side relay is required to deliver the same bit (from a particular source side relay) to both sinks, then the relay simply copies the obtained bit on both links to the sinks. If the relay is required to supply two different bits to the two sinks, the matchings are to different bits, the relay fetches the two bits from the respective source side relays on two different links (as per matching), and supplies them to the two sinks via two different links. This matching on an as-needed basis minimizes link usage. But every time a new sink is added, new flows should be initiated to make all bits available to the new sink. Can we {\em prepare} the network to be in a state of readiness so that upon addition of a new sink, it is merely the new sink that does the necessary work to obtain all bits?

Our next goal is to modify Algorithm \textsf{\small MaxFlow} into one that pushes two steps and then pulls, as in Algorithm \textsf{\small ALLCAST}, yielding a decentralized algorithm that easily extends to the case of multiple sinks.

Consider the single source single sink case again, and the following algorithm.

\vspace*{.1in}

\begin{itemize}[{\bf Algorithm} \textsf{\small MaxFlowPUSHPULL}:]
  \item {\em Push step 1}: The source node $s$ floods $np(1-\varepsilon)$ links with one bit per link. We shall call the bits $b_1, b_2, \ldots, b_{np(1-\varepsilon)}$ and the recipient nodes of these bits as the owners $O_1, O_2, \ldots, O_{np(1-\varepsilon)}$ of the respective bits. All other nodes are termed relays and indexed $R_{np(1-\varepsilon)+1}, \ldots, R_n$.
  \item {\em Push step 2}: Each owner $O_i$ pushes his bit $b_i$ one more level, but only to neighbors who are not owners, and to the sink $t$ if there is a link to the sink. Owner-owner links are unutilized.
  \item {\em Pull step 1}: The sink $t$ collects all bits sent directly by owners.
  \item {\em Pull step 2}: The sink $t$ identifies the list of additional bits needed, the list of relays it is connected to, the list of bits they have in their possession, and does an appropriate matching of relays with the required bits. It then pulls the desired bits from these relays via the by now all-too-familiar matching. $\hfill \IEEEQEDopen$
\end{itemize}

The bit-map for this setting is much simpler (see Table \ref{tb:maxflow-bitmap}). The columns are indexed by the bits. The rows are indexed by the nodes, with the first $np(1-\varepsilon)$ representing the owners and the rest representing the relays. Row $i$, when it corresponds to owner $O_i$ (which is when $1 \leq i \leq np(1-\varepsilon)$) has a 1 only on the $i$th column. But when row $i$ corresponds to a relay (which is when $i > np(1-\varepsilon)$), it has entry $X_{ij} = 1$ if $O_j$ is connected to $R_i$. Clearly, the presence or absence of this link is independent of the status of all other links, and $X_{i,j}$ is a Bernoulli$(p)$ random variable, when $ i > np(1-\varepsilon) \geq j$.

\vspace*{.1in}
\begin{table}[htb]
\begin{center}
\caption{Bit-map for one source one sink flow}
\label{tb:maxflow-bitmap}
\noindent {\small
\begin{tabular}{|c|cccc|} \hline \hline
                              & $b_1$   & $b_2$       & $\cdots$ & $b_{np(1-\varepsilon)}$ \\ \hline \hline
$O_1$             & 1     & 0         & $\cdots$ & 0 \\
$O_2$             & 0     & 1         & $\cdots$ & 0 \\
$\vdots$                      &       &           & $\ddots$ &   \\
$O_{n p (1-\varepsilon)}$   & 0     & 0         & $\cdots$ & 1 \\ \hline \hline
$R_{n p (1-\varepsilon)+1}$       &       &           &          &   \\
$\vdots$                                &       &           & $((X_{i,j}))$&   \\
$R_n$                             &       &           &          &   \\ \hline \hline
\end{tabular}
}
\end{center}
\end{table}

\vspace*{.1in}

We then have the following result.


\begin{thm}
  \label{thm:maxflow}
  For any $\varepsilon > 0$, the following event occurs almost surely: for all but finitely many $n$, the algorithm \textsf{\small MaxFlowPUSHPULL} succeeds in transporting all $np(1-\varepsilon)$ bits from the source $s$ to the sink $t$. $\hfill \IEEEQEDopen$
\end{thm}


\begin{IEEEproof}
This is almost immediate. If the algorithm fails, one of the following must happen.

(1) The event $A^{(n)}_1$ occurs, which is the event that node $s$ is connected to less than $n p (1-\varepsilon)$ relays. By Lemma \ref{lem:binomial-concentration}, there is a $c_1 > 0$ such that for all sufficiently large $n$, we have $\Pr \{ A_1^{(n)} \} \leq e^{-n c_1}$.

(2) The event $A_2^{(n)}$ occurs, which is the event that the sink $t$ is connected to a number of owners outside the range $n p (1-\varepsilon) \cdot p (1 \pm \varepsilon)$. Again by Lemma \ref{lem:binomial-concentration}, there is a $c_2 > 0$ such that for all sufficiently large $n$, we have $\Pr \{ A_2^{(n)} \} \leq e^{-n c_2}$ for some $c_2 > 0$.

(3) The event $A_3^{(n)}$ occurs, which is the event that the sink $t$ is connected to fewer than $\beta_n := n (1 - p (1 - \varepsilon)) \cdot p (1 - \varepsilon)$ relays. Again by Lemma \ref{lem:binomial-concentration}, there is a $c_3 > 0$ such that for all sufficiently large $n$, we have $\Pr \{ A_3^{(n)} \} \leq e^{-n c_3}$.

(4) If $A^{(n)}_1 \cup A^{(n)}_2 \cup A^{(n)}_3$ does not occur, the number of bits that remain to be pulled is at least $ n p (1-\varepsilon) - n p(1-\varepsilon) \cdot p (1 + \varepsilon)$ which is at most $\beta_n$. The number relays that can help the sink pull these bits is at least $\beta_n$. For the algorithm to fail, the event $M^{(n)}$, that there is no coverage of the yet-to-be-pulled bits by the available relays with each relay accounting for at most one bit (capacity constraint), must then occur. This implies that if a particular set of $\beta_n$ relays are chosen, there is no coverage of the required bits. This further implies that any superset of $\beta_n$ bits that includes the yet-to-be-pulled bits cannot be covered by the $\beta_n$ chosen and available relays.

The matrix rows corresponding to the $\beta_n$ chosen relays (rows) and the $\beta_n$ chosen bits (columns) is a $\beta_n \times \beta_n$ square submatrix whose entries are conditionally iid Bernoulli$(p)$ random variables. Again, we may view this as a bipartite graph with the chosen relays on the one side and chosen bit indices on the other side. Thus, if $A^{(n)}_1 \cup A^{(n)}_2 \cup A^{(n)}_3$ does not occur, but $M^{(n)}$ does, then there is no matching on the random bipartite graph. Using Theorem \ref{thm:bipartite-matching-n}, the probability that such a matching does not exist, conditioned on $(A^{(n)}_1 \cup A^{(n)}_2 \cup A^{(n)}_3)^c$, is upper bounded by $\gamma(\beta_n)$.

Thus, the event that the sink is unable to pull all the bits implies the event
\[
  A_1^{(n)} \cup A^{(n)}_2 \cup A^{(n)}_3 \cup M^{(n)},
\]
and its probability is upper bounded by
\begin{equation}
  \label{eqn:prob-failure}
  e^{-n c_1} + e^{-n c_2} + e^{-n c_3} + \gamma(\beta_n).
\end{equation}
This is summable by Lemma \ref{lem:summability}, and the rest follows.
\end{IEEEproof}


Instead of one sink, suppose we have two sinks $t_1$ and $t_2$ that are not connected directly to each other or directly to the source. The source has to transport all its $np(1-\varepsilon)$ bits to each of the two sinks using only the $n$ relay nodes. We may continue to use \textsf{\small MaxFlowPUSHPULL} with the following extension. The two push steps are common. But each sink simply executes its own pull steps based on the connections it sees at its end and the information from its helper nodes. Using the union bound, it immediately follows that Theorem \ref{thm:maxflow} holds for one source and two sinks when there are no direct connections between the set of nodes constituted by the source and the sinks.

\begin{figure}[tb]
\centering
\includegraphics[width=3in, height=2.4in]{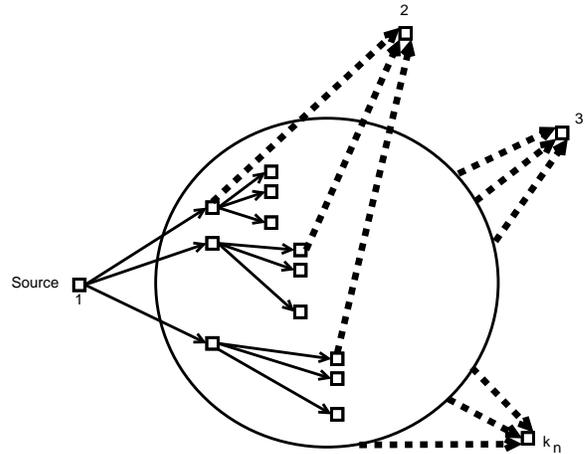}
\caption{The \textsf{\small relay}$(k_n,n)$ network. Source pushes bits to owners who then push to relays (solid lines). The sinks pull the bits from either owners or relays (dashed lines).}
\label{fig:push-pull}
\end{figure}

Indeed, we can say something much stronger. One version that suffices to address the multicast setting of the next section is the following. Consider a scenario where there is one source $s$ and a total of $k_n-1$ sinks $t_1, t_2, \ldots, t_{k_n-1}$ where $\sup_{n \geq 1} \frac{k_n}{n} \leq C$ for some $C < \infty$. The source and the sinks have no links among themselves, but are connected through a network of $n$ relays. See Figure \ref{fig:push-pull}. The internal links between the relays and the links between the source/sinks and the relays are iid Bernoulli($p$) random variables. The source wishes to transfer all its bits of information to each of the sinks. Let us denote this random network as \textsf{\small relay}$(k_n, n)$.

\begin{thm}
  \label{thm:maxflow-relay}
  For any $\varepsilon > 0$, the following event occurs almost surely: for all but finitely many $n$, the algorithm \textsf{\small MaxFlowPUSHPULL}, with the pull stages implemented by each sink, succeeds in transporting all $np(1-\varepsilon)$ bits from the source $s$ to each of the $k_n-1$ sinks on the \textsf{\small relay}$(k_n,n)$ network. $\hfill \IEEEQEDopen$
\end{thm}


\begin{IEEEproof}
Observe that the first three terms in the upper bound for the probability of failure in (\ref{eqn:prob-failure}) decay exponentially fast in $n$. The last term $\gamma(\beta_n)$ satisfies $\sum_{n \geq 1} n \gamma(\beta_n) < \infty$. Since there are $k_n - 1 = O(n)$ sinks, by the union bound, the probability that the algorithm fails for some sinks is at most $Cn \left( e^{-n c_1} + e^{-n c_2} + e^{-n c_3} + \gamma(\beta_n) \right)$. This upper bound is summable, and the rest follows.
\end{IEEEproof}

A related model was considered by Ramamoorthy et al. \cite{200508TIT_RamShiWes}. In their random network model, between each pair of nodes, there are two links, one in each direction, with equal but random capacity. The random variables were again iid. They identified how the minimum cut capacity, which is also the multicast capacity in directed settings, scales with the number of relays. Our achievability result is, in contrast to that of \cite{200508TIT_RamShiWes}, constructive. Further, thanks to the undirected nature of links in our model, our ability to choose directions flexibly enables us to reach the network upper bound, asymptotically, with flows.

\section{Multicast: Achievability}
\label{sec:multicast-achievability}

We now return to the setting of $n$ nodes of which $k_n$ are in a multicast session. Node 1 is the source node and nodes $2, 3, \ldots, k_n$ are the sinks. Our goal in this section is to show that the upper bound (\ref{eqn:eta-n-kn-limit-bound}) is achievable. While one could in principle proceed as in Catlin et al. \cite{Catlin-et-al} to prove achievability, we shall directly jump to a constructive proof.


\begin{thm}
  \label{thm:multicast}
  For the multicast problem with $k_n$ nodes in the session, let $\lim_{n\rightarrow\infty} k_n /n = \alpha \in [0,1]$. We then have
  \[
    \lim_{n \rightarrow \infty} \frac{\pi_n(k_n)}{n} = \left( 1 - \frac{\alpha}{2} \right) \mathbb{E}[C] \quad \mbox{ a.s.}
\]   $ \hfill \IEEEQEDopen$
\end{thm}


\begin{IEEEproof}
As in the proof of Theorem \ref{thm:allcast}, converse was already shown in (\ref{eqn:eta-n-kn-limit-bound}). So showing achievability suffices, and further showing it on Erd\H{o}s-R\'{e}nyi random graphs with parameter $p$ suffices. Moreover, as before, it is enough to show that: {\em For any $\varepsilon > 0$, the following event occurs almost surely: for all but finitely many $n$, there is an algorithm that succeeds in transporting $\pi_n(k_n) \geq n \left( 1 - \alpha/2 \right) p (1 - 2\varepsilon)$ bits from the source to each of the $k_n - 1$ sinks.} We claim that this holds.

We first dispose two easy cases.

When $\alpha = 0$, this follows from Theorem \ref{thm:maxflow-relay}, by simply ignoring the links between the session nodes and by using \textsf{\small MaxFlowPUSHPULL} and $n(1-\varepsilon)$ relays, and with pulls implemented at each of the sink nodes.

When $\alpha = 1$, pretend that all nodes are in session and implement \textsf{\small ALLCAST}. The result follows from Theorem \ref{thm:allcast-algo-asym}.

Only the case when $0 < \alpha < 1$ remains, for which we will use a combination of the above.

Observe that the subset of session nodes alone form a complete graph with $k_n$ vertices for which Theorem \ref{thm:allcast-algo-asym} is applicable. Using \textsf{\small ALLCAST} and without using any of the relay nodes, we have that the source can distribute
\begin{equation}
  \label{eqn:session-only}
  \pi_n^{(1)} \geq \frac{k_n}{2} p (1 - \varepsilon)
\end{equation}
bits to the other $k_n - 1$ nodes in the session, for all but finitely many $n$, almost surely. (Summability of the probability upper bound sequence holds since $k_n = \Omega(n)$).

Removing these direct links between the session nodes, we end up with the graph in Figure \ref{fig:push-pull}, where the session nodes are now only connected to the $m_n = n - k_n$ relay nodes. The link to each relay node from each session node has Bernoulli$(p)$ capacity. Further the relay nodes have interrelay link capacities that are independent Bernoulli$(p)$ random variables. By Theorem \ref{thm:maxflow-relay}, using \textsf{\small MaxFlowPUSHPULL}, the source can distribute
\begin{equation}
  \label{eqn:relay-only}
  \pi^{(2)}_n \geq m_n p (1 - \varepsilon)
\end{equation}
bits to the $k_n-1$ sinks (solely with the help of the relay nodes), for all but finitely many $n$, almost surely. (Summability of the probability upper bound sequence holds since $m_n = \Omega(n)$).

The result immediately follows from (\ref{eqn:session-only}) and (\ref{eqn:relay-only}) since $\pi(k_n) \geq \pi_n^{(1)} + \pi_n^{(2)}$ and $k_n/2 + m_n = n - k_n/2 \geq n(1 - \alpha/2)(1-\varepsilon)$ for all sufficiently large $n$.
\end{IEEEproof}

\section{Vanishing Link Probabilities}
\label{sec:vanishing-pn}

Our results extend to the case when $p$ is a function of $n$, denoted $p_n$, and vanishes but sufficiently slowly. We shall focus only on the allcast problem. The results for multicast can be obtained in an analogous fashion.


\begin{thm}
  Let $p_n = \sqrt{\frac{\tau_n \log n}{n}}$ where $\tau_n \rightarrow \infty$ but $p_n \rightarrow 0$. For any $\varepsilon > 0$, the following event occurs almost surely: for all but finitely many $n$, the algorithm \textsf{\small ALLCAST} succeeds in distributing $\frac{1}{2} n p_n(1-\varepsilon)$ bits to each of the $n$ nodes. Furthermore, $\lim_{n \rightarrow \infty} \frac{\pi_n}{n p_n} = \frac{1}{2}$ almost surely. $\hfill \IEEEQEDopen$
\end{thm}


\begin{IEEEproof}
The proof of the first part is similar to the proof of Theorem \ref{thm:allcast-algo-asym}, with some additional effort to get better probability upper bound estimates. Again, we argue that the probability that algorithm \textsf{\small ALLCAST} fails is summable over $n$. If the algorithm fails for a particular $n$, at least one of the following events must have occurred.

1) The event $A_1^{(n)}$ occurs, which is defined to be the event that there are fewer than $\frac{1}{2}np_n(1-\varepsilon)$ vertices connected to node 0. By Lemma \ref{lem:binomial-concentration}, applied with $q = p_n/2$, there is some $c_1 > 0$ such that for all sufficiently large $n$, we have
      \[
        \Pr \{ A_1^{(n)} \} \leq e^{-n \cdot \frac{1}{2} p_n \cdot \varepsilon^2 / 3} = e^{ - c_1 \sqrt{n \tau_n \log n} }.
      \]

2) For some node $t$, the event $A^{(n)}_2(t)$ occurs, which is defined to be the event that node $t$ is connected to a certain number of owners outside the range $\frac{1}{2}np_n(1-\varepsilon) \cdot \frac{1}{2} p_n (1 \pm \varepsilon)$ with links pointing towards $t$. (The case when node $t$ is an owner leads to one fewer number of owners which as before is absorbed into $(1 \pm \varepsilon)$ factor). Again by Lemma \ref{lem:binomial-concentration}, there is some $c_2 > 0$ such that for all sufficiently large $n$, we have
      \begin{eqnarray}
        \Pr\{ A^{(n)}_2(t) \} & \leq & e^{- \frac{1}{2} n p_n (1 - \varepsilon) \cdot \frac{1}{2} p_n \cdot \varepsilon^2 / 3 } \nonumber \\
        & \leq & e^{- c_2 n p_n^2 } \nonumber \\
        & = & e^{- c_2 \tau_n \log n} ~ = ~ \frac{1}{n^{c_2 \tau_n}}. \label{eqn:owner-bound}
      \end{eqnarray}
      Note that $c_2$ can be arbitrarily small because of the $\varepsilon^2$ factor. Since we need $n$ times $\Pr\{ A^{(n)}_2(t) \}$ to go to zero, see (\ref{eqn:n-times-proberror}) which comes later, it is here where we utilize the assumption that $\tau_n \rightarrow \infty$.

3) Let $A_1^{(n)}$ not occur. Then there are exactly $\frac{1}{2}np_n(1-\varepsilon)$ owners. For some node $t$, the event $A^{(n)}_3(t)$ occurs, which is the event that the node $t$ is connected to fewer than
  \begin{eqnarray}
    \beta_n & := & \left( n - \frac{1}{2}np_n(1-\varepsilon) \right) \cdot \frac{1}{2} p_n (1-\varepsilon) \nonumber \\
            & = & \frac{1}{2}np_n(1-\varepsilon) \cdot \left( 1- \frac{1}{2}p_n(1-\varepsilon) \right) \label{eqn:beta-n}
  \end{eqnarray}
  relays with links pointing towards $t$. (As before, the case of 1 less relay when node $t$ is a relay is easily handled). Once again by Lemma \ref{lem:binomial-concentration}, there is a $c_3 > 0$ such that for all sufficiently large $n$, we have
  \begin{eqnarray*}
    \Pr\{ A^{(n)}_3(t) ~|~ ( A_1^{(n)} )^c \} & \leq & e^{- \left( n - \frac{1}{2}np_n(1-\varepsilon) \right) \cdot \frac{1}{2}p_n \cdot \varepsilon^2 / 3 } \\
    & \leq & e^{ - c_3 \sqrt{ n \tau_n \log n} }.
  \end{eqnarray*}

4) For some node $t$, if $A_1^{(n)} \cup A^{(n)}_2(t) \cup A^{(n)}_3(t)$ does not occur, then the event $M^{(n)}(t)$ occurs, which is the event that node $t$ is unable to pull the desired bits. We claim that
      \begin{equation}
        \label{eqn:no-matching-pn}
        \Pr \left\{ M^{(n)}(t) ~|~ \left( A_1^{(n)} \cup A^{(n)}_2(t) \cup A^{(n)}_3(t) \right)^c \right\} \leq \delta_n
      \end{equation}
      where
      \begin{equation}
        \label{eqn:no-matching-summable-pn}
        \sum_{n=1}^{\infty} n \delta_n < \infty.
      \end{equation}

The event that the algorithm fails is thus a subset of
\[
  A_1^{(n)} \bigcup_{t=1}^n \left( A^{(n)}_2(t) \cup A^{(n)}_3(t) \cup M^{(n)}(t)\right)
\]
whose probability is upper bounded via the union bound and (\ref{eqn:no-matching-pn}) by
\begin{equation}
   \label{eqn:n-times-proberror}
   n \cdot \left( e^{-c_1 \sqrt{n \tau_n \log n}} + \frac{1}{n^{c_2 \tau_n}} + e^{- c_3 \sqrt{n \tau_n \log n}} + \delta_n \right).
\end{equation}
By (\ref{eqn:no-matching-summable-pn}) and the assumption that $\tau_n \rightarrow \infty$, we see that this bound is summable.

What remains is to prove (\ref{eqn:no-matching-pn}) and (\ref{eqn:no-matching-summable-pn}).

As before, the probability on the left-hand side of (\ref{eqn:no-matching-pn}) is upper bounded by the probability that there is no matching in a bipartite graph with $\beta_n$ vertices and link probability $p_n$.

We first sharpen Lemma \ref{lem:Bollobas-consequence2}. The bound in (\ref{eqn:Fa-bound}), after noting that we now have $\beta_n$ vertices on one side, can be sharpened (see \cite[p.174]{Bollobas}) to
\begin{eqnarray*}
    \Pr \{ F_a \} \leq 2 \binom{\beta_n}{a} \binom{\beta_n}{a-1} (1-p_n)^{a(\beta_n-a+1)} \\
    \cdot \left( \binom{a(a-1)}{2a-2} \cdot p_n^{2a - 2} \right)
\end{eqnarray*}
where the extra term within parentheses in the second line can be included because it is an upper bound (via the union bound) on the probability that some $2a-2$ links, among the possible $a(a-1)$ links from $A$ to $\Gamma(A)$, are active. Recall that $a$ is an integer satisfying $2 \leq a \leq (\beta_n + 1)/2$. Using the bounds $\binom{m}{a} \leq \left( \frac{em}{a} \right)^a$ and $(1 - x) \leq e^{-x}$, we get
\begin{eqnarray*}
  \lefteqn{ \Pr \{ F_a \} } \\
    & \leq &  2 \left( \frac{e \beta_n}{a} \right)^a \left( \frac{e \beta_n}{a-1} \right)^{a-1} \left( \frac{ea}{2} \right)^{2a-2} p_n^{2a - 2}  \\
      & & \quad \quad \quad \cdot \left( e^{- \beta_n p_n a \left( 1-\frac{a}{\beta_n}+\frac{1}{\beta_n} \right) } \right) \\
  & \leq & 2 \left( \frac{e \beta_n}{a} \right) \left( \frac{e^2 \beta_n p_n}{2} \right)^{2a-2} \left( 1 + \frac{1}{a-1} \right)^{a-1} \\
      & & \quad \quad \quad \cdot \left( e^{- \beta_n p_n a \left( 1-\frac{a}{\beta_n}+\frac{1}{\beta_n} \right) } \right) \\
  & \leq & C n p_n \left( \frac{e^2 n p_n^2 }{4} \right)^{2a-2} e^{-a n p_n^2 (1 - 2 \varepsilon) / 4}
\end{eqnarray*}
for some finite constant $C$, where in the last inequality we have used $(1 + 1/k)^k \leq e$, the bound $1 - (a-1)/\beta_n \geq 1/2$ when $2 \leq a \leq (\beta_n + 1)/2$, and the obvious upper and lower bounds on $\beta_n$ from (\ref{eqn:beta-n}). Now, using $np_n^2 = \tau_n \log n$, we get
\begin{eqnarray*}
  \Pr \{ F_a \}
  & \hspace*{-2mm} \leq \hspace*{-2mm} & C \sqrt{n \tau_n \log n} \left( \frac{e^2 \tau_n \log n }{4} \right)^{2a-2} n^{-a \tau_n (1 - 2 \varepsilon) / 4} \\
  & \hspace*{-2mm} \leq \hspace*{-2mm} & C \left( 16 \frac{\sqrt{n}}{e^4(\tau_n \log n)^{1.5}} \right) \left( \frac{ e^4 (\tau_n \log n)^2 }{16 n^{\tau_n(1-2\varepsilon)/4}} \right)^a.
\end{eqnarray*}
Since the term inside the second parentheses converges to zero as $n \rightarrow \infty$, it follows that for all sufficiently large $n$ and some finite constants $C_1$ and $C_2$, we have
\begin{eqnarray*}
  \sum_{a = 2}^{(\beta_n+1)/2} \Pr \{ F_a \}
  & \hspace*{-2mm} \leq \hspace*{-2mm} & C_1 \left( \frac{\sqrt{n}}{(\tau_n \log n)^{1.5}} \right) \left( \frac{ e^4 (\tau_n \log n)^2 }{16 n^{\tau_n(1-2\varepsilon)/4}} \right)^2 \\
  & \hspace*{-2mm} = \hspace*{-2mm} & C_2 \left( \frac{\sqrt{n} (\tau_n \log n)^{2.5}}{n^{\tau_n(1-2\varepsilon)/2}} \right) \\
  & \hspace*{-2mm} =: \hspace*{-2mm} & \kappa_n.
\end{eqnarray*}
The probability that there is no matching is then upper bounded by $\delta_n := \kappa_n + 2 \beta_n (1 - p_n)^{\beta_n}$. The second term is upper bounded, using the bounds on $\beta_n$, as
\[
  2 \beta_n (1 - p_n)^{\beta_n} \leq n p_n e^{- n p_n^2 (1 - 2\varepsilon) / 2} = \frac{\sqrt{n \tau_n \log n}}{ n^{\tau_n (1 - 2\varepsilon)/2 }}.
\]
From these two bounds, using $\tau_n \rightarrow \infty$, it is clear that not only $\delta_n \rightarrow 0$, but in addition, $\sum_{n \geq 1} n \delta_n < \infty$. This establishes (\ref{eqn:no-matching-pn}) and (\ref{eqn:no-matching-summable-pn}) and proves validity of algorithm \textsf{\small ALLCAST}.

The above achievability result also establishes that $$\liminf_{n \rightarrow \infty} \frac{\pi_n}{np_n} \geq \frac{1}{2}.$$ The upper bound $$\limsup_{n \rightarrow \infty} \frac{\eta_n}{np_n} \leq \frac{1}{2}$$ follows from (\ref{eqn:eta-n-bound}) and Lemma \ref{lem:binomial-concentration}. This concludes the proof of the second statement.
\end{IEEEproof}

The extension to multicasting can be done similarly.

\section{Discussion}
\label{sec:discussion}

We began with the problem of allcast and multicast capacity region for multiple allcast and multiple multicast. Yet, we largely focused on single allcast or single multicast with just one sender and with remaining nodes of the session as receivers. But study of single multicast suffices, thanks to the result \cite[Cor. 4.a]{Li-et-al} of Li et al. on transferability of rates across sources (even with network coding). It is therefore clear how the established results imply the validity of (\ref{eqn:all-cast-region}) and (\ref{eqn:multicast-region}). The requirement that the session nodes be identical for each of the multiple multicasts is crucial for this transferability.

Moreover, we largely studied multicasting techniques that do not use network coding. One message coming out of this work is that though network coding provides a coding advantage in specific undirected scenarios, and one such example can be found in Li et al. \cite{Li-Li-Lau-2006}, in large dense random undirected networks of the variety studied in our paper the coding advantage is at most $1 + o(1)$ in the number of nodes. While our results applied to graphs $G(n, p_n)$ with $p_n \rightarrow 0$, we did require that $p_n$ vanishes sufficiently slowly. In particular, $p_n = \sqrt{(\tau_n \log n)/n}$ so that a typical node has degree $n p_n = \sqrt{n \tau_n \log n}$. These are well connected, but by no means sparse graphs. This naturally raises two questions. (1) Can one extend these results to some useful classes of sparse random graphs? (2) Can one find the rate at which the expected rates for the proposed strategies converge to their asymptotic limits, and show concentration around the expectations?

The result of asymptotically negligible network coding advantage in single or multiple multicast settings (with identical session nodes) may evoke the question of a possible connection with a conjecture of Li and Li \cite{Li-Li-2004} for multiple unicasts. Li and Li \cite{Li-Li-2004} conjectured that for multiple unicast, network coding provides no coding advantage in undirected graphs. While their conjecture holds true for some specific classes of undirected graphs (\cite{Harvey-et-al}, \cite{200606TIT_JaiEtAl}), the general conjecture remains unresolved. The negligible gain for multicasting in random graphs studied here arises from the dense interconnectivity between relays. The bottlenecks are primarily at the periphery\footnote{This is also why the simplest of partitions yields asymptotically tight upper bounds in Theorem \ref{thm:upperbound}.}. So there does not seem to be much insight that one can glean from our study to prove or disprove the Li and Li conjecture for multiple unicasts in undirected networks.

While we studied multiple multicasts, our communication application naturally restricted us to a single set of session nodes. We thus had to study Steiner tree packings for a single subset of nodes. VLSI applications require efficient packing of Steiner trees across a multiplicity of such subsets (or nets; see \cite{199702MP_GroMarWei}). One could apply our random network framework to such problems and attempt to devise similar quick-but-dirty algorithms. This is an interesting topic that is beyond the scope of this paper.

\section*{Acknowledgements}
The authors would like to thank Prof.\ Navin Kashyap for bringing references \cite{Barahona} and \cite{Catlin-et-al} to their attention.




\begin{IEEEbiographynophoto}{Vasuki~Narasimha~Swamy} received the B.Tech. degree in electrical engineering from the Indian Institute of Technology Madras in 2012. She is currently pursuing her Ph.D. degree at the University of California at Berkeley.
\end{IEEEbiographynophoto}

\vspace*{-.2in}

\begin{IEEEbiographynophoto}{Srikrishna~Bhashyam}(S'96--M'02--SM'08) received the B.Tech. degree in electronics and communication engineering from the Indian Institute of Technology (IIT), Madras, India, in 1996, and the M.S. and Ph.D. degrees in electrical and computer engineering from Rice University, Houston, TX, in 1998 and 2001, respectively. From June 2001 to March 2003, he was a Senior Engineer at Qualcomm CDMA Technologies, Campbell, CA. Since May 2003, he has been with the Department of Electrical Engineering, IIT Madras, where he is currently an Associate Professor. His current research interests are in resource allocation, adaptive transmission, code design, and information theory for multiterminal wireless communication systems.
\end{IEEEbiographynophoto}

\vspace*{-.2in}

\begin{IEEEbiographynophoto}{Rajesh~Sundaresan}(S'96--M'00--SM'06) received the B.Tech. degree in electronics and communication from the Indian Institute of Technology Madras, the M.A. and Ph.D. degrees in electrical engineering from Princeton University in 1996 and 1999, respectively. From 1999 to 2005, he worked at Qualcomm Inc. on the design of communication algorithms for wireless modems. Since 2005, he has been with the Department of Electrical Communication Engineering, Indian Institute of Science, Bangalore. His interests are in the areas of communication networks and information theory. He is an associate editor of the {\sc IEEE TRANSACTIONS ON INFORMATION THEORY} for the period 2012-2015.
\end{IEEEbiographynophoto}

\vspace*{-.2in}

\begin{IEEEbiographynophoto}{Pramod~Viswanath}(SM'10--F'13) received the Ph.D. degree in EECS from the University of California at Berkeley in 2000. He was a member of technical staff at Flarion Technologies until August 2001 before joining the ECE department at the University of Illinois, Urbana-Champaign. He is a recipient of the Xerox Award for Faculty Research from the College of Engineering at UIUC (2010), the Eliahu Jury Award from the EECS department of UC Berkeley (2000), the Bernard Friedman Award from the Mathematics department of UC Berkeley (2000), and the NSF CAREER Award (2003). He was an associate editor of the {\sc IEEE TRANSACTIONS ON INFORMATION THEORY} for the period 2006–2008.
\end{IEEEbiographynophoto}

\end{document}